      [1999/12/01 v1.4c Il Nuovo Cimento]
\documentclass{cimento}
\usepackage{epsfig,amssymb,txfonts}

\title{Discovery of the optical afterglow of XRF\,040812: VLT and $Chandra$ observations}
\author{P.~D'Avanzo\from{ins:x}\from{ins:y}\ETC,
D.~Malesani\from{ins:z},
S.~Campana\from{ins:x},
S.~Covino\from{ins:x}\\
A.~Moretti\from{ins:x}
G.~Tagliaferri\from{ins:x}
        \atque
G.~Chincarini\from{ins:x}\from{ins:k}}

\instlist{\inst{ins:x} INAF, Osservatorio Astronomico di Brera, via Bianchi 46, I-23807 
       Merate (Lc), Italy
  \inst{ins:y} Universit\`a degli studi dell'Insubria, Dipartimento di Fisica e Matematica, via Valleggio 11, I-22100 
                          Como, Italy
  \inst{ins:z} International School for Advanced Studies (SISSA/ISAS) via Beirut 2-4, I-34016 
                          Trieste, Italy
  \inst{ins:k} Universit\`a degli studi di Milano Bicocca, Piazza delle Scienze 2, I-20100 
                          Milano, Italy}
\PACSes{\PACSit{98.70.Rz}{$\gamma$-ray sources; $\gamma$-ray bursts}
\PACSit{01.30.Cc}{Conference proceedings}
}
\begin{document}

\maketitle

\begin{abstract}
We present $Chandra$ and VLT observations of the X$-$Ray Flash XRF\,040812. The X$-$ray analysis reveals with high precision the
position of a hard, fading source. A careful analysis of our $I-$band VLT images taken starting 17 hours after the burst 
led to the discovery of the optical afterglow superimposed to a bright ($I=21.5$) host galaxy. The optical afterglow is seen 
decaying with an index of 1.1. We do not detect any jet break and supernova rebrightening in the optical light curve. 
The bright apparent luminosity of the host galaxy allows us to get a rough estimate of the redshift, comparing with a 
set of GRB/XRF host galaxies with known luminosity and redshift. Such comparison suggests a redshift of XRF\,040812 in 
the range $0.3 \leq z \leq 0.7$. This is also consistent with the lack of emission features in our spectrum. The low inferred 
redshift is in agreement with the idea that XRFs are low-luminosity, closer events.
\end{abstract}

\section{Observations and discovery of the afterglow}

The X-Ray Flash XRF 040812 was discovered by the INTEGRAL satellite on 2004 Aug 12.251 UT, with the IBIS/ISGRI instrument 
in the 15-200 keV band~\cite{ref:G}. Our optical observations at the VLT could start only 0.7 days after the XRF. In order to overcame 
any saturation problem due to the presence of a very bright star (V$\sim$11) we placed an occulting bar on the detector even 
if this meant to hide part of the INTEGRAL error box. Two epochs of Chandra observations (5 and 10 d after the burst) 
revealed the presence of a fading, uncatalogued source within the INTEGRAL error box~\cite{ref:CM1}. 

Following on the Chandra results, we re-analized our VLT data discovering that the position of the Chandra candidate fell 
just behind the occulted area of the detector in both epochs. Luckily, thanks to the dithering applied to the images, in 
two out of ten exposures taken for each epoch the position of the Chandra candidate was visible. This enabled us to discover 
an optical source coincident with the position of the X-ray afterglow. Comparing the two epochs we found that this source 
exhibited significant fading, confirming it as the optical afterglow transient. We continued monitoring the afterglow with 
three other sets of images until ~220 days after the burst; during this monitoring the source continued its decay until it 
reached the level of the host galaxy. This analysis confirms the identification of the Chandra candidate as the afterglow of 
XRF 040812. 

\section{Optical results}

The decay is well described by a power law with slope $\alpha = 1.1 \pm 0.6$ (F(t) = F$_0$t$^{-\alpha}$). This value is consistent with the 
X-ray decay index. Within the photometric errors, no jet-break or supernova-like rebrightening is apparent in our data. 
The host galaxy magnitude, after correcting for the high Galactic interstellar absorption, is I = 19.97 $\pm$ 0.12. We note 
that our luminosity value for the host galaxy of XRF 040812 is about 1 mag brighter than reported by~\cite{ref:SOD}.


An optical spectrum of the HG was also taken in the range 4500-8600 Å but, due to the low S/N, no emission lines were 
detected. However, since the spectra of GRB host galaxies usually show strong emission lines, the lack of the most prominent 
of them can be used to constrain the redshift range. For example, the non-detection of the H${\alpha}$ and O III 5006 lines would 
require z$\geq$0.3 and z$\geq$0.7 respectively. 

We can use the apparent luminosity itself to get a rough estimate of the distance to the galaxy. To this extent, from a set 
of~\cite{ref:W}, we extracted all the host galaxies with a known $I$ magnitude and we have added to this 
sample the I magnitudes of host galaxies with known redshift discovered during the Swift era. All the magnitudes are corrected 
for interstellar absorption. A relatively low redshift, z $\leq$ 0.7, is suggested for the host galaxy of XRF 040812. This limit, 
combined with our spectroscopic results, suggests a redshift of XRF 040812 in the range 0.3 $\leq$ z $\leq$ 0.7. We have also extracted 
a set of 156 galaxies from the SDSS catalogue in the magnitude range $I_{host} \pm 0.3$ mag and found that the mean redshift of this 
sample is z=0.43 $\pm$ 0.17~\cite{ref:PDA}.

\begin{figure}
\includegraphics[height=.16\textheight]{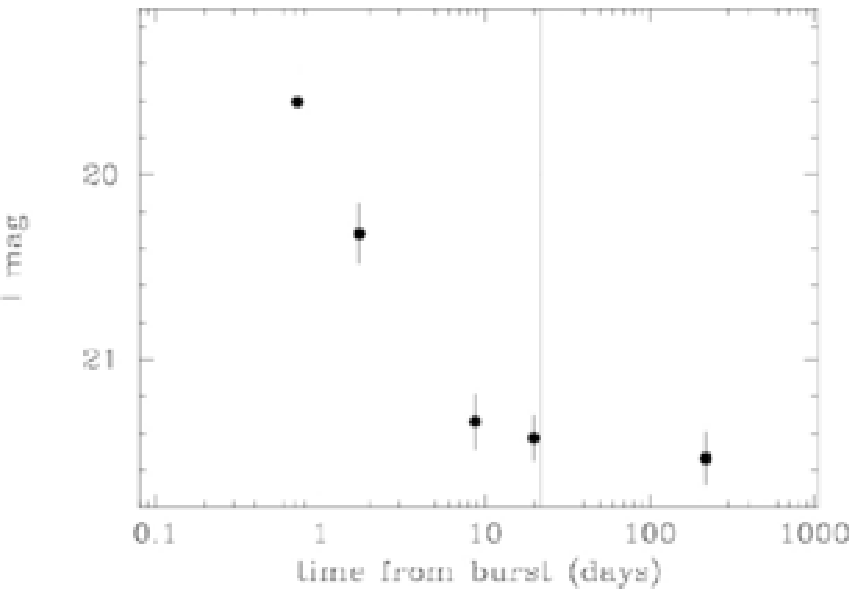}
\includegraphics[height=.16\textheight]{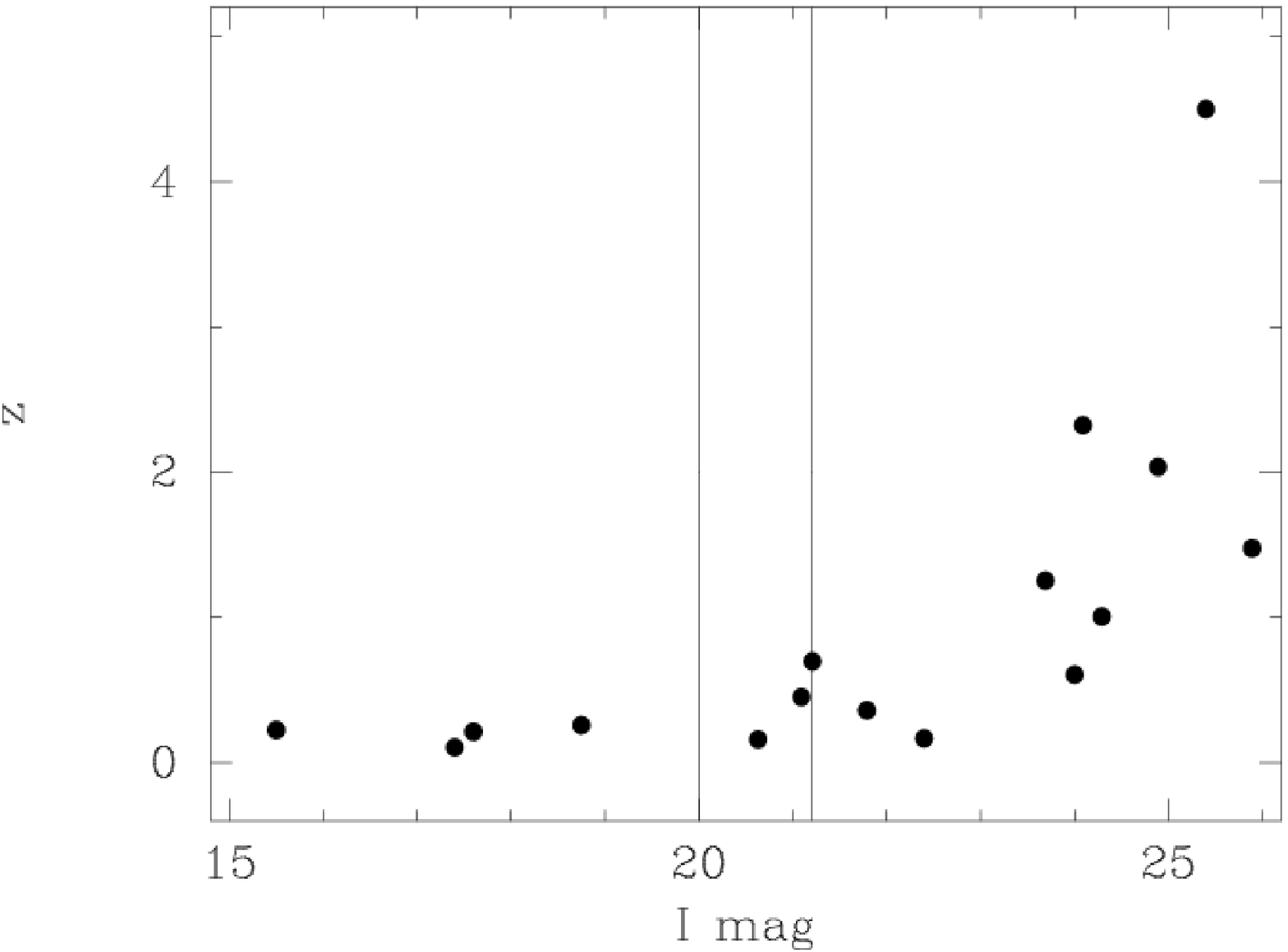}
\includegraphics[height=.16\textheight]{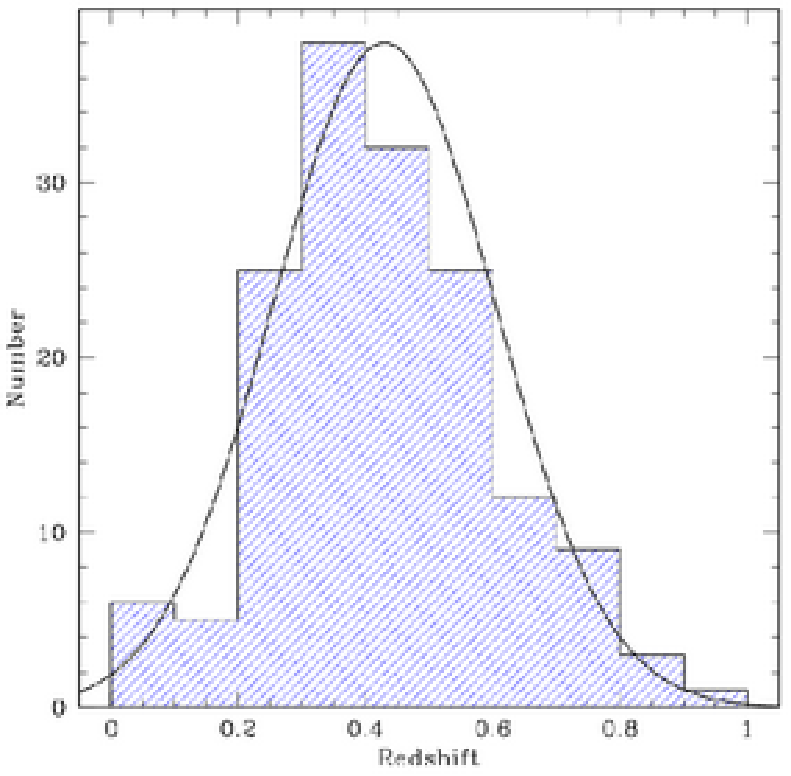}
\caption{({\it From left to right}). Light curve of the afterglow of XRF\,040812; the solid line indicates the epoch 
of our spectrum. Redshift distribution a f a sample GRB/XRF host galaxies; the solid lines indicates the $I$
magnitude (corrected for the Milky-Way absorption) for the HG of XRF\,040812. Redshift distribution of a set of
galaxies of the SDSS catalogue in the magnitude range $I_{HG\,040812} \pm 0.3$ mag.} 
\end{figure}

\end{document}